\documentclass[sigplan,authorversion,nonacm]{acmart}

\usepackage{xspace}

\AtBeginDocument{%
  }

\setcopyright{acmcopyright}
\copyrightyear{2018}
\acmYear{2018}
\acmDOI{XXXXXXX.XXXXXXX}

\acmConference[Conference acronym 'XX]{Make sure to enter the correct
  conference title from your rights confirmation emai}{June 03--05,
  2018}{Woodstock, NY}
\acmPrice{15.00}
\acmISBN{978-1-4503-XXXX-X/18/06}

\newcommand{\ie}{\emph{i.e.,}\xspace}
\newcommand{\eg}{\emph{e.g.,}\xspace}
\newcommand{\etc}{\emph{etc.}\xspace}
\newcommand{\etal}{\emph{et al.}\xspace}

\newcommand{\seclabel}[1]{\label{sec:#1}}

\newcommand{\lstlabel}[1]{lst:#1}

\newcommand{\lstref}[1]{Listing~\ref{lst:#1}}
\newcommand{\secref}[1]{Section~\ref{sec:#1}}

\usepackage{xcolor}
\usepackage{textcomp}
\usepackage{listings}
\definecolor{source}{gray}{0.9}
\lstdefinelanguage{sysmel}{
	basicstyle=\ttfamily\small,
	comment=[l]{\#\#},
	morecomment=[s]{\#*}{*\#},
	commentstyle=\color{purple}\ttfamily
}

\begin{document}

\title{The Design and Implementation of an Extensible System Meta-Programming Language}

\author{Ronie Salgado}
\email{ronie@desromech.cl}
\orcid{1234-5678-9012}
\affiliation{%
  \institution{Desromech EIRL - AOne Games SpA}
  \city{Santiago}
  \country{Chile}
}

\begin{abstract}
System programming languages are typically compiled in a linear pipeline
process, which is a completely opaque and isolated to end-users.
This limits the possibilities of performing meta-programming in the
same language and environment, and the extensibility of the compiler
itself by end-users. We propose a novel redefinition of the compilation process
in terms of interpreting the program definition as a script. This evaluation is
performed in an environment where the full compilation pipeline is implemented
and exposed to the user via a meta-object protocol, which forms the basis for
a meta-circular definition and implementation of the programming language
itself. We demonstrate the feasibility of this approach by bootstrapping a
self-compiling implementation of Sysmel, a static and dynamic typed Smalltalk
and C++ inspired programming language.

\end{abstract}



\maketitle

\section{Introduction}
\paragraph{System vs Non-System Language} An important dichotomy in the
classification of programming languages is on whether a programming language is
meant for low-level close to the machine \emph{System} programming or not. System
programming languages such as C and C++ tend to have semantics with a direct
translation towards unoptimized machine operations. These semantics allows a
programmer using these language having a mental model. This cognitive mental
model facilitates learning and debugging activities \cite{canas1994mental}. It
allows system programmer to have direct control of the machine, which facilitates
writing high-performance code by avoiding unneeded abstraction
layers such as having bytecode interpreter, JIT or garbage collector that
introduces latency and non-determinism in execution times.

On the other hand, non-system programming languages such as Java, C\# and Python, are
languages that do not have a direct correspondence with machine operations.
These non-system programming language
facilitate the software development activity by providing abstractions such as
automatic memory management, faster iteration cycles via interpretation,
dynamic and duck typing, \etc. The presence of these abstractions increase
the runtime cost of the program execution, and
they also sacrifice the capability of having this close to the metal mental
model. However, these abstraction are desirable because they improve software
development productity, and they are used when execution performance can be
sacrificed.

\paragraph{Language Impedance Mismatch} In multiple application
domains, the simultaneous usage of a system and a non-system programming
language is required. A high performance critical core is written in the
low-level system programming language. The UI and non-critical performance
sections are commonly written in higher-level languages which are typically used
for scripting purposes. This also facilitates the extensibility of an application
by people without a programming expertise, like an end user of a scriptable
application such as a spreadsheet editor (\eg VBA in Microsoft Excel\cite{walkenbach2010excel}).

The usage of at least two completely different programming languages is a common
practice in the videogame industry. Commercial game programming is an activity where high-performance and
productivity is important \cite{sweeney2006next}, and striving for a balance
between them is a necessity. Game engines such as Unreal Engine\cite{unrealengine}
and Unity\cite{haas2014history}, typically have a high performance core written
in C++, and a high-level language like C\# or Blueprint used for scripting and
game design. Using multiple languages facilitates productivity
in terms of reducing game testing and design iteration cycles by programming and
non-programming people. However, the connection between two completely different
languages such as C++ and the Blueprint, the visual scripting language used by Unreal,
requires the maintenance or generation of wrapper code. These wrappers are
typically maintained by hand or generated by a specialized offline tool that
imposes restriction on the programming language features that can be used. They
are some some general purposes tools like SWIG \cite{beazley1996swig} for
solving this problem, but their usage might be precluded by project specific
constraints.

\paragraph{Fixed Compilation Pipeline} The fixed compilation pipeline of C and
C++ does not provide extension points in the language compiler itself. Accessing
to the compiler data structures in an scriptable way inside might be an ideal
mechanism for generating custom application specific reflection metadata
required for supporting garbage collection, and automatic scripting language connection.
Extensible compilation also facilitate metaprogramming, and the construction
of DSL embedded directly in the host language \cite{renggli2010embedding}.

\paragraph{Unified Programming Language and Environment} We propose the design
and construction of a programming language that can be used simultaneously in
both context. We propose using this language as a script that defines how to build a
program, whose execution constructs a \emph{Program Entity Graph}. Different
subgraphs can be obtained by tracing a subset of the program entities from an
user specified root set. In the case of system programming, this root set
is typically composed of only the main entry point. For a fully reflective
environment, where the language is not used for system-programming, the root set
is composed of the main entry and the global namespace object. By changing
the set of program entities traced, we can compile down or up different features
of the programming language, which facilitates adapting its usage for system and
non-system program.

In \secref{sysmel-language} we describe the design of Sysmel, a Smalltalk, Lisp
and C++ inspired System Metaprogramming Language. In \secref{bootstrapping} we
describe our bootstrapping processs along with the challenges faced by it.

\section{Sysmel language}\seclabel{sysmel-language}
\subsection{Design}

\paragraph{Influences} In this section we describe the design and
implementation of Sysmel, a System Metaprogramming Language, with a design
inspired mostly in Smalltalk, Lisp and C/C++. With the objective of unifying
system and non-system programming, with an extensible compiler and
metaprogramming facilities we take strong inspiration on these three important historical
languages: 1) from Lisp, we take important the concepts of macros as function from AST
into AST \cite{steele1996evolution}, meta-circular evaluation \cite{reynolds1972definitional}, and the
Meta Object Protocol used for defining an Object Oriented Programming environment \cite{kiczales1991art};
2) from Smalltalk, we take object-oriented programming via message passing, blocks as closures,
the importance of a minimalistic syntax, and reflection as a mechanism for
meta-circular definition; 3) and from C/C++, we take primitive types, pointers
and direct memory access. We infer a concept of using static primitives types
as a mechanism for definining translational. These type defined semantics
facilitate a direct to machine operation mental-model.

\paragraph{Design Wishlist} From these influences we distill the following
feature wishlist that we want to support in a single programming language, even
if there are conflicting between them:

\begin{enumerate}
\item Minimalistic convenient to use flexible syntax.
\item Everything should \emph{look} like an object at the syntax level.
\item Everything must be typed. Dynamic typing is realized via static typing.
\item Type inference support. The minimum is supporting the local type
    inference like the C++ \emph{auto} keyword. Stronger type inferences
    algorithms like Hindley-Milner \cite{hindley1969principal}\cite{milner1978theory}
    are desirable, but not required.
\item Block closures.
\item Arbitrary compile time evaluation support.
\item Lisp style macros which are functions from AST into AST functions.
\item Primitive types which are directly supported by the target machine.
\item Pointers and direct memory accesses.
\item Manual memory management support.
\item Optional automatic memory management via garbage collection support.
\item Compile time and optional runtime reflection.
\item The ability for extending and modifying all of the compilation pipeline
	stages.
\end{enumerate}

\paragraph{Syntax in a postcard} The Sysmel syntax is strongly based on the
syntax from Smalltalk, but there are additions and changes taken from C/C++
to facilitate supporting different programming styles. Sysmel syntax
is minimalistic and it is based around the concepts of message sending,
function application, and the construction of commonly used data structures.
Everything is an expression and returns a value with a
specific type. See \lstref{sysmel-syntax} for a postcard that illustrate
the whole Sysmel base syntax. Higher-level syntactical constructs are realized
via composing these base syntax elements, and by using metaprogramming
techniques that manipulate the AST during compilation time.

\begin{figure}[htb]
\centering
\lstset{language=sysmel,caption={Sysmel syntax postcard},label=\lstlabel{sysmel-syntax}}
\begin{lstlisting}[frame=single]
## Literals
0 . -1 . 2r1101_0011 . 16rFF1F_F2F3.
2.5 . -3.5e-2.
#hello . #with:with: . #+ . ## Symbols
"Hello World\n\r".          ## Strings
'A' . '\''.                 ## Characters
#(1 2 test (2.5 3))         ## Literal array

1, 2, 3.                               ## Tuple
#{first: 1. #second : 2 . "third": 3}. ## Dictionary
#[1u8 . (2 + 3) asUInt8].              ## Byte array

## Identifiers
UInt8 . true . false . nil . void .
(RawTuple::new:) . RawTuple::slotAt:put: .

## Message sends
2 negated.             ## Unary
2 + 3 * 5.             ## Binary
Array with: 1 with: 2. ## Keyword
2 ::+ 3.               ## Low precedence binary
a := 2.                ## Assignment
with: #x with: 42.     ## Keyword without receiver
Int32[5sz].            ## #"[]:" message
doSomething{5sz}.      ## #"{}:" message
doSomething#[1u8].     ## #"#[]:"  message

## Message chain
(Array new: 3sz)
    at: 0sz put: 1;
    at: 1sz put: 2;
    yourself.

malloc(16sz). ## Function application

## Lexical block
{ let: #x with: 2 . x }.

## Block closure.
{:(Int32)x :y :: Int32 | x + y} (1i32, 3 i32)

## AST Node quote.
`'a.

## AST Node quasi-quote
``(a `,unarySelectorNode (`@callArgumentNodes)).

\end{lstlisting}
\end{figure}

\subsection{Semantic Analysis Meta-Object Protocol}

\paragraph{AST Protocol} The scanning and parsing
stages can be done by using traditional approaches such as manually written recursive
descent parsing, or by using more extensible approaches like parser combinators.
Parsing produces nodes which conform to the Meta Object Protocol. The top level
nodes respond to the \emph{\#analyzeAndEvaluateWithEnvironment:} message. This single message is used for
simultaneous analysis and evaluation of the top-level parsed source code. The
environment received by it is used for looking identifier values. Lambda
nodes are evaluated into closure objects, which are composed of two parts: a
capture vector, and a function definition object which contains analyzed arguments and body 
nodes. The analysis of the elements of a function definition are performed by
sending the \emph{\#analyzeWithEnvironment:} message onto the argument, result
type and body nodes. This message is responsible of returning a newly analyzed
node where its value type is solved, and its children nodes are also analyzed
recursively. Once a function definition is analyzed, it can be evaluated via two
different mechanisms: 1) direct interpretation of the analyzed node, supported
through the \emph{\#evaluateWithEnvironment:} message. 2) compilation into a
bytecode or an IR that can be further optimized and compiled into machine code.
We support these two mechanisms. In summary, the AST node semantic analysis MOP
is composed of the following messages: \emph{\#analyzeAndEvaluateWithEnvironment:},
\emph{\#analyzeWithEnvironment:}, \emph{\#evaluateWithEnvironment:},
\emph{\#compileBytecodesDirectlyWith:} and \emph{\#generateSSAValueWith:}.

\paragraph{Extensible AST} AST nodes are defined as ordinary class instances inside of Sysmel. New AST
nodes can be defined by just subclassing from ASTNode and then overriding the
required methods. New AST nodes can be exposed in the language through macros.
In fact, the local variable definition AST node is not present in the sysmel
language syntax, but we expose it through two different mechanism: 1) macros
like \emph{\#let:with:} and \emph{\#let:type:with}; and 2) the \emph{let}
metabuilder (See \secref{metabuilders}).

\paragraph{Function Application Analysis} Function applications are analyzed in
two phases: first as an unexpanded function application, where the functional
object can be a macro. The macro is invoked with the non-analyzed parameter
nodes as arguments. The node returned by the macro is analyzed recursively. In the
case of expanded function applications, the analysis of the application node is
delegated onto the functional object type. This allows treating any object or
value as a functional object. In the case of ordinary functions, its
evaluation is performed by constructing an activation environment with the
evaluated arguments. In other cases, the \emph{\#applyWithArguments:} message
is sent to the functional object. One important optimization is always performed
if possible. We define functionally \emph{pure functions} in terms of
\emph{observable external side effects}, so we allow programmers to perform internal
definitions through impure imperative mechanisms. For this reason any function
can be marked as \emph{pure} function. A pure function application that only
uses literal nodes is \emph{always evaluated in compile time}, and the application node
is replaced by a literal node with the evaluation result. This application of
referential transparency for pure functions is mandatory, and we use it for
constructing derived type literals.

\paragraph{Message Send Analysis} In the case of message sends, there are also
multiple analysis phases. First, the receiver node is analyzed, and the actual
message send analysis is delegated onto the receiver node type. The receiver
type first analyses the message selector. If the analyzed selector is a literal,
then the corresponding macro or method is looked up through multiple
dictionaries. If the found method is a macro, then it is expanded by receiving the
receiver and argument nodes as parameters. If the method is not a macro, then
there are two cases: if the method does not require dynamic dispatch (\ie it
cannot be overriden by a subclass), then the message send node is converted into
a function application node which is analyzed recursively. If dynamic dispatch
is required, then the remaining analysis of the message send is delegated onto
the method type. If no method was found, if the receiver is not a dynamic type
a semantic error is raised, otherwise a generic analysis is performed for the
arguments of dynamic message sends.

\paragraph{Type System Protocol} The type system is another important side of the MOP. Types
are also objects, and they are in fact instances of the class \emph{Type}. The analysis
of some AST nodes is delegated into specific types. This facilitates defining
type specific translational semantics, binding message sends to
methods statically, and defining type specific macros. We also use the type system
for constructing pointer and reference types. C++ style references are used for 
constructing mutable variables. A reference type is constructed by
responding to almost no message, except for \emph{\#:=} used for assignments, and
\emph{address} used for converting a reference into pointer. Pointers can
be converted into references through the \emph{\_} message. With these messages
we support the semantics of the C pointer operators (\&, *). The type system
MOP is much larger than the MOP used for AST nodes. The following is a non-exhaustive
list of some messages that are part of the MOP exposed in Type: \emph{\#methodDictionary}, 
\emph{\#lookupSelector:}, \emph{\#analyzeAndEvaluateMessageSendNode:forReceiver:withEnvironment:}, \emph{\#analyzeMessageSendNode:withEnvironment:},
\emph{\#analyzeAndTypeCheckFunctionApplicationNode:withEnvironment:},
\emph{\#analyzeAndTypeCheckSolvedMessageSendNode:withEnvironment:}, \emph{\#ref}, \emph{\#pointer}.

\subsection{Metabuilders}\seclabel{metabuilders}
The \emph{Builder} pattern in Smalltalk is a common pattern for the construction
of objects through successive message sends in a chain. We extend this pattern
onto the meta-level by defining the concept of a metabuilder. A metabuilder is a
builder that operates on syntactic language elements, and they can be seen as
statefull macros. Metabuilders are instanced by invoking a macro function or
macro identifier known as the metabuilder factory. Metabuilders are ordinary
objects where their classes override the \emph{\#analyzeAndEvaluateMessageSendNode:forReceiver:withEnvironment:}
and \emph{\#analyzeMessageSendNode:withEnvironment:} methods by delegating them
onto the metabuilder instance. This delegation is always possible for the
simultaneous analysis and evaluation case, and it is only possible if the
metabuilder instance is present on a literal node for the AST analysis case. We use
metabuilder for making higher-level syntactic looking constructs which look
familiar to C++ and Java programmers. We also use them for hiding the actual
manipulation of the underlying program entities which are also constructed
through ordinary message sends. See \lstref{sysmel-metabuilder-usage} for
an example on how code can look like a different language when using metabuilders, even
though the base syntax from \lstref{sysmel-syntax} is still the same.

\begin{figure}[htb]
\centering
\lstset{language=sysmel,caption={Metabuilder Usage},label=\lstlabel{sysmel-metabuilder-usage}}
\begin{lstlisting}[frame=single]
public class SampleClass
  superclass: Object; definition: {
    public field first => Int32.
    
    public method add: (x: Int32) ::=> Int32
        := first + x.
}.

function sampleFunction(x: Int32, y: Int32) => Int32
    := SampleClass new first: x; add: + y.

printLine(sampleFunction(2i32, 3i32).
\end{lstlisting}
\end{figure}

\subsection{Optimization and code generation pipeline}
\paragraph{High-Level IR} The optimization and code generation pipeline unlike the semantic analysis is a
much more traditional process. We perform code generation and optimization
successive translation from the analyzed AST into different intermediate
representations (IR). We first translate the AST into a high-level SSA based IR
with a design inspired by LLVM \cite{lattner2004llvm}, where we model our base
language operation semantics like function applications, message sends, local
variable allocation (alloca), pointer and object slot load and stores. At this
level we represent primitive type intrinsics as function calls. In our current
implementation we perform some optimizations like constant propagation,
control flow simplification and inlining. We are planning on having many more
optimizations at this level.

\paragraph{Middle-Level IR} The high-level SSA form is translated into a mostly
portable middle-level three address code IR which is also in SSA. The instructions
of this IR is composed by tuples that contain a single machine primitive
operation and its operands. We use this IR for performing lower-level
optimizations like combining comparison with branches, and register allocation.
We also compute the stack frame layout computation, and some phases of debug
information generation during this stage before generating the next stage which
is assembly.

\paragraph{Low-Level IR} Our low-level IR is assembly code. We use this IR for
final machine code generation, and also for generating object files with
included debug information. A subset of the program object graph is serialized
into the data segment of the resulting object file, and references between
objects are annotated with the required relocations. We are capable of
generating platform specific relocatable ELF, COFF and Mach-O object file.
Since we do not implement a linker, we rely on the standard linker provided by
each operating system for constructing actual standalone executable programs.

\section{Bootstrapping}\seclabel{bootstrapping}
\paragraph{Minimal C Implementation} Due to the circularity on the language
definition, performing a proper bootstrap is a tricky and complicated process.
We have attempted multiple times to construct and bootstrap
this system. In our current implementation, we perform a phase0 compilation by
constructing a minimal implementation in C. This minimal implementation takes
care of parsing, base language semantic
analysis, it uses the LISP2 compacting garbage collection algorithm \cite{cohen1983comparison}.
To reduce bootstrapping development iteration cycles we implemented a register based
bytecode, and a simplistic x86\_64 JIT. The bootstrap environment uses a logical
object-model where raw native pointers are not exposed. In this object model we
have three kind of objects: immediates (encoded as tagged pointers),
byte tuples, and pointer tuples. All of the objects can be seen as a tuple that
contains a type (another tuple), an identity hash value, and the size of the
tuple itself in bytes. With this simplistic object model we construct a complete
Smalltalk style image environment in Sysmel. The base objects and types are
defined by hand in C, the intrinsic operations are exposed as functions which
have the name of the primitive annotated. We also implemented a minimal bootstrap parser
and semantic analyzer in C.

\paragraph{Metastability Problems} When writing the actual semantic analyzer of
Sysmel in Sysmel we had to be extra careful. Each time a new MOP method is
implemented, its new definition is immediately being used for subsequent
semantic analysis. Running onto different meta-stability
issues when performing this kind of definitions is a big issue. We solved these
problems by introducing the new definitions on a specific order, and by also
anotatting the new method as functions that require an eager semantic analysis
before they are installed as the new semantic analyzed and executable method.

\paragraph{Self Feeding AST and Program Graph} The traditional compiler
bootstrapping process is performed via feeding the source code representation
to subsequent versions of the compiler. In our case, we are compiling from the
already analyzed in memory AST which is stored as part of the program entiy
graph. The required program entities are traced from the global namespace object,
and the \emph{main} entry point function. The analyzed AST is used for constructing
the SSA based IR and subsequent lower-level IR before generating machine code.
The program entity object graph is serialized onto the data section of the object
files, and references are annotated with relocations.

\section{Limitations}\seclabel{limitations}

\paragraph{Frontend not validated by bootstrap} One important limitation of our
bootstrapping approach is that it only validates the quality of our middle-end
and backend. The frontend of our compiler is not being validated by the
bootstrapping process. Unlike a traditional bootstrap that starts clean from
source-code, we start from already analyzed in memory AST nodes. We are skipping
completely the source code parsing and semantic analysis stages during
bootstrapping after the first phase. For this reason the frontend implementation
is not being validated by the bootstrap.

\paragraph{Memory usage} Our bootstrapping process serializes a copy of the
fully semantic analyzed AST and metaobjects that compose the program entity
definitions. This incurs on a larger memory usage since a lot of
compilation-only metadata is kept by the process. However, this same metadata
might be used to construct development and debugging tools.

\section{Related Work}\seclabel{related-work}

\paragraph{Bootstrapping Reflective Systems} Polito \etal describe the
complexities and the metastability issues that happen when bootstrapping a
highly reflective system like Pharo \cite{polito2014bootstrapping, polito2015bootstrapping}.

\paragraph{Embedding Languages in Pharo} Helvetia by Renggli is \etal
\cite{renggli2010embedding} is a framework for embedding languages inside of
Pharo. We take inspiration inspiration from Helvetia for multiple elements such
as the quasi-quoting of preceding by an extra backquote the standard Common Lisp
operators. Our monadic parser framework is based on PetitParser by Kurs \etal
\cite{kurs2013petitparser}, another component used by Helvetia.

\paragraph{Bee Smalltalk} Bee is a Smalltalk implementation which is also
completely defined in itself. Bee is also capable of constructing native
executable through similar reflection based serialization process. Instead of
relying on supporting C style primitive types for constructing the base runtime,
the usey different mechanisms which they call ``undermethods, underprimitives
and inline nativization of bytecodes.'' \cite{pimas2014design}


\section{Conclusions and future work}\seclabel{conclusions}

We described the central concepts behind the metacircular
definition of Sysmel, a programming language designed for system and non-system
programming. We also proved the feasibility for constructing this system by
bootstrapping an self-compiling version of Sysmel capable of compiling and
optimizing itself through three full self-compilation cycles.

In the future, we would like to continue improving our compilation and
optimization infrastructure. We would like to perform benchmarks with a much
more optimized version of Sysmel, on realistic applications to further validate
the language. In this venue, it would also be desirable to validate the usage
of Sysmel with people.

\bibliographystyle{ACM-Reference-Format}
\bibliography{references}

\end{document}